\documentclass{elsart}
\usepackage{epsfig}
\begin{document}
\begin{frontmatter}
\title{Deuterium burning in Jupiter interior}
\author[cauni,cainfn]{M. Coraddu},
\author[cainfn,cauni]{M. Lissia},
\author[cauni,cainfn]{G. Mezzorani},
\author[cainfn,to]{P. Quarati}
\address[cauni]{Physics Dept., Univ. Cagliari, I-09042 Monserrato, Italy}
\address[cainfn]{I.N.F.N. Cagliari, I-09042 Monserrato, Italy}
\address[to]{Phys. Dept. and INFM, Politecnico Torino,
                 I-10125 Torino, Italy}
\date{1 September 2001}
\begin{abstract}
We show that moderate deviations from the Maxwell-Boltzmann energy
distribution can increase deuterium reaction rates enough to contribute to
the heating of Jupiter. These deviations are compatible with the violation
of extensivity expected from temperature and density conditions inside
Jupiter.
\end{abstract}
\end{frontmatter}
\section{Introduction}
Jupiter emits more radiation then it recives from the sun: the origin of
this excess heat is still uncertain and debated. Possible explanations are:
release of gravitational potential energy due to the planet contraction 
and/or Helium sedimentation \cite{Guillot:1999bw},
decaying of radioactive isotopes in the core
\cite{Hu:80}, or deuterium burning \cite{Ho:91}. Each of these hypotheses
has difficulties \cite{Ou:98}; in particular, standard calculations of
deuterium burning reaction rates predict negligible contribution to the planet
thermal balance, in spite of the substantial enhancement due to electron and
ion screening effects \cite{Ho:91,Ic:93}.

In a strongly coupled plasma, anomalous diffusion and time correlation
effects originate non-Maxwellian two-body relative energy distribution
that can be parameterized with and, in same cased, assume the same functional
form that appears in the contest of Tsallis non-extensive thermodynamics
\cite{Kaniadakis:1997,Kaniadakis:1998my}.
As demonstrated for the solar core, small changes of the tail
of the energy distribution can strongly modify the fusion rates
without affecting mechanical properties (hydrostatic equilibrium and the
sound speed) that depend on the mean value of the distribution
\cite{Coraddu:1999yb}.

Since the internal conditions of Jupiter indicate the existence of a strongly
coupled plasma, we investigate the effects of the consequent small
deviations from the standard Maxwell-Boltzmann (MB) statistics on deuterium
burning rates and the possibility that deuterium burning could play, or 
have played in the past, a role in Jupiter thermal balance.
\section{Jupiter interior and standard deuterium burning}
Jupiter interior is a mixture of liquid metallic hydrogen and helium,
with density and temperature within the ranges:
$\rho_J = 2 - 5 $ g cm$^{-3}$ and $kT = 1-2 $ eV; during
planet formation the central temperature should have been of the order of
$10-20$ eV \cite{Ou:98}. If we assume the reference values
$\rho_J = 5 $ g cm$^{-3}$ and $kT = 2 $ eV, the corresponding
density of H, D, He and electrons are: 
$n_p = 2.4\times 10^{24}$\/ cm$^{-3}$, $n_D = 3 \times 10^{-5} n_p$,
$n_{He} = 6.25\times 10^{-2} n_p$, and $n_e = 1.125\, n_p$.

Being the electron Fermy energy, $E_F \sim 50$ eV , much greater than the
thermal energy, the electron gas is fully degenerated. The electron and
ion plasma parameters, $\Gamma_{j}  = (Z_i Z_j e^2)/(a^2 kT) $,
where $a=\left(3Z/(4\pi n)\right)^{1/3}$ is the Wigner-Seitz radius, 
are: $\Gamma_e \approx \Gamma_i \approx 16 $. Therefore, the interior
of Jupiter is a Strong Coupled Degenerated Plasma.

Jupiter excess energy flux $\Phi_J  = 5.4 \times 10^{3}$ erg cm$^{-2}$
sec$^{-1}$ implies  an excess luminosity:
$L_J =  3.5 \times 10^{24}$ erg sec$^{-1}$. Each D(p,$\gamma$)$^3$He reaction
produces an energy $Q_{pD} = 5.493$ MeV. Therefore, deuterium burning
is relevant to Jupiter thermal balance only if the rate is greater than a
threshold $r_t =
(\rho_J L_J ) / (M_J Q_{pD} ) \approx 1 \textrm{sec}^{-1} \textrm{cm}^{-3}$,
where we have used as Jupiter mass and radius: $M_J = 1.90 \times 10^{33}$ g
and $R_J  = 7.14 \times 10^{9}$ cm.

Two-body reaction rates in a thermal plasma can be expressed as
$ (1 + \delta_{ij}) r_{i,j} = n_i n_j \langle \sigma v \rangle_M $,
where $\langle \sigma v \rangle_M$ is the thermal average of the reaction
cross section $\sigma$ times the relative velocity $v$, and $n_i$ is the
number density of species $i$.

Cross sections between charged particles below threshold are dominated by
the penetration factor $e^{-b / \sqrt{E}}$, where 
$b = \pi \sqrt{2 \mu} Z_i Z_j e^2 / \hbar$ and  $\mu$\/ is the particle
reduced mass. In fact the astrophysical factor
$ S(E) =  \sigma(E) E \exp{(b\sqrt{E})}$ has a mild dependence on $E$ in
absence of resonances.

There exist no experimental determinations of $S(E)$\/ for deuterium reactions
at energies in the eV range. Theoretically motivated low-energy extrapolations
can increase the deuterium \cite{An:99} are: $S(0) = 2.0\times 10^{-7}$
MeV~b for D(p, $\gamma$)$^3$He, $S(0) = 5.0\times 10^{-2}$MeV~b for 
D(d, n)$^3$He and $S(0) = 5.6\times 10^{-2}$ MeV~b for D(d, p)T.

If the thermal distribution is Maxwellian, $\langle \sigma v \rangle_M
\sim\int_0^\infty E\sigma(E) e^{-E/kT} dE$, thermonuclear rates are
dominated by the product of the exponentials $e^{-E/kT}$ (MB distribution)
and $e^{-b/\sqrt{E}}$ (penetration factor), product  that has its maximum
at the Gamow Peak Energy, $E_0$. In the weak screening regime 
$E_s < E_0$\/ we can use a saddle point expansion around $E_0$ and find:
\begin{equation}
       r_{i,j} = f_e f_i \frac{n_i n_j}{1 + \delta_{ij}}
                 \sqrt{\frac{2}{\mu}}\, \frac{S(E_0)}{(kT)^{3/2}}
                 I_{\mathrm{max}} \Delta 
         \label{eq:MaxScreenReac}
\end{equation} 
where $E_0 =  \left( b kT / 2 \right)^{2/3}$ is the  Gamow Energy,
$I_{\mathrm{max}} = \exp\left( -3 E_0 / kT \right)$ is the integrand at
$E=E_0$, and $ \Delta = 4 \left(E_0  kT /3 \right)^{1/2}$ measures the width
of the Gamow Peak.
The two correction factors, $f_e = \left( 1 - E_s/E_0 \right)
           \exp\left(E_s / kT\right)$ for electrons and
$ f_i =  \exp\left(H(0) /kT \right)$ for ions, take care of screening.

In fact screening is very important in Jupiter interior.
Electron screening lowers the coulomb barrier of $E_s = e^2/D_s $, where
$D_s$\/ is the short-range screening distance, which is determinated
through the incipient Rudberg method \cite{Ic:93}.
Many body correlations modify also ion wave functions in plasmas: it is
possible to define an equivalent mean field potential $H(r)$\/ which
is related to the plasma parameter $\Gamma$ in the contest of Ion Sphere
Model: $H(0) / kT = 1.057 \Gamma_s $
where $\Gamma_s\; $\/ is the screened ion plasma factor
$ \Gamma_s = \Gamma_i \exp\left( - a /D_s \right)$ and $\Gamma_i $\/
is the ion plasma factor.

Despite the large enhancement factors ($f_e\sim 10^{11}$, $f_i\sim 10^2$)
deuterium burning remains negligible: $r_{pD}\sim 10^{-19}$ and
$r_{DD}\sim 10^{-26}$\/ sec$^{-1}$\/ cm$^{-3}$. Rates are much lower than
the threshold rate $r_t$ for any relevant temperature.
\section{Non-Maxwellian deuterium burning}
Non extensive Thermodynamics, introduced by Tsallis \cite{Ts:88} has been
applied to many different fields. In particular, it has been observed
\cite{Kaniadakis:1997,Kaniadakis:1998my}
that small deviations from Maxwellian distribution
can be described by weakly non-extensive Tsallis distributions, and that such
deviations can have dramatic consequences for nuclear reaction rates in the
solar core~\cite{Coraddu:1999yb}.

We can calculate the effects of non-extensivity on nuclear reaction rates by
substituting the Maxwell-Boltzmann distribution, $\exp{(-E/kT)}$, with
the Tsallis one, $\left[1 - (1-q)E/kT \right]^{q/(1-q)}$, inside the
thermal average:
\begin{equation}
\langle \sigma v \rangle_M \sim
   \int_0^\infty e^{ -\frac{b}{\sqrt{E+E_s}}}
        \left(1 - (1-q)\frac{E}{kT}  \right)^{q/(1-q)}
          S(E + E_s) \d E \, .
 \label{eq:TsInt}
\end{equation}
The limit $q\to 1$ recovers the Maxwellian case.
Analogously to the Maxwellian calculation, we find a $q$-dependent
Gamow Peak Energy   $E_{0q}$, and, in the weak screening condition
$E_s < E_{0q}$, we can compute again the reaction rates at first order
in the saddle point expansion:
\begin{equation}
 r_{i,j,q} = \frac{n_i n_j}{1 + \delta_{ij}}
              \sqrt{\frac{2}{\mu}}
          \left( 1 - \frac{E_s}{E_{0q}} \right) 
           \frac{S(E_{0q})}{(kT)^{3/2}} I_{q,\mathrm{max}} \Delta_q \, ,
    \label{eq:TsScreenReac}
\end{equation}
where now $E_{0q} = E_0 \xi^2(T,q) / q^{2/3}$ is the $q$-dependent Gamow
Peak Energy, 
\[
I_{q,\mathrm{max}} = 
     \exp\left[ -\frac{b}{\sqrt{E_{0q}+E_s}} + \frac{q}{1-q}
        \ln\left( 1 - (1-q)\frac{E_{0q}}{kT}  \right) \right] 
\]   is the integrand at $E=E_{0q}$, and
\[
 \Delta_q = \Delta  \left( \xi(T,q) / q^{1/3} \right)^{5/2}
        \left(
         1+ \frac{ q(1-q) 2E_0 /(3kT) \xi^5(T,q)/q^{5/3} }
                 {\left( 1-(1-q) (E_{0q}-E_s ) / kT\right)^2}
             \right)^{-1/2}
\] measures the peak width. The other terms are: 
$a = (1-q) b^{2/3} / \left(4q^2 kT \right)^{1/3}$,
$b= 1+ (1-q) E_s / kT $, and
\[
3 \xi(T,q) = - a +  2^{-1/3}
           \left[\left(27b-2a^3-c\right)^{\frac{1}{3}}+
                 \left(27b-2a^3+c\right)^{\frac{1}{3}} \right]
\] with $c=\sqrt{27b(27b-4a^3)}$, while $E_0$ and $\Delta$ are the
same as in the MB case.

Note that it is not possible to factorize an electron enhancement factor
as for the MB distribution.

\section{Discussion and conclusions}
\begin{figure}[hp]
\begin{center}
\epsfig{bbllx=49pt,bblly=197pt,bburx=540pt,bbury=600pt,%
file=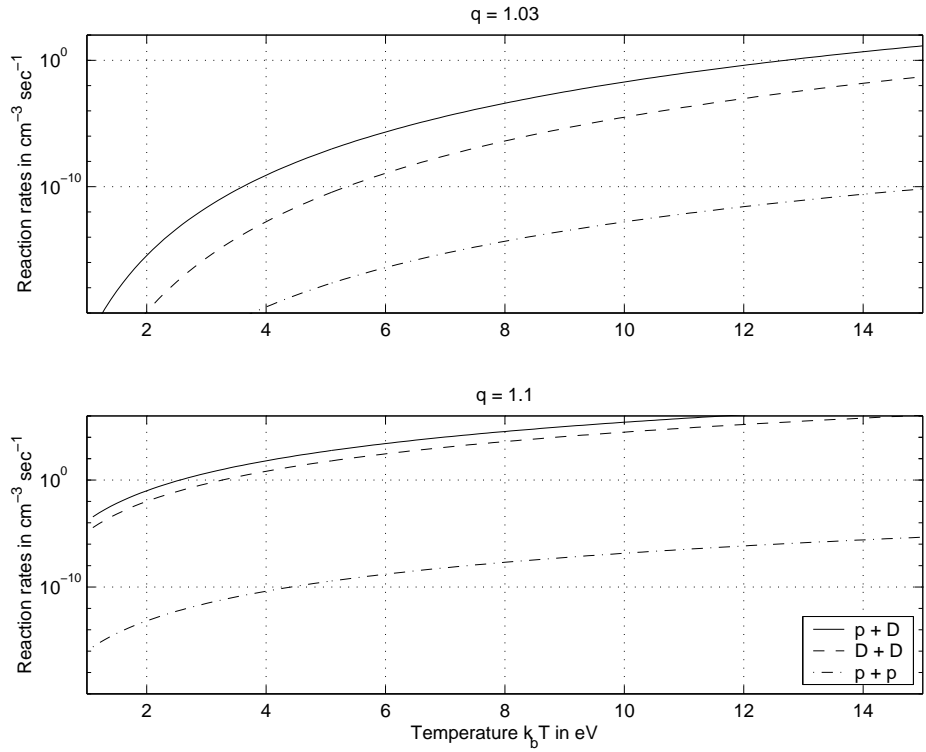,height=10cm}
\end{center}
%\vspace{-5mm}
\caption[aa]{Reaction rates for the reactions p+D (solid), D+D (dashed) and
p+p (dot-dashed) as function of temperature for two values of the 
Tsallis parameter: $q=1.03$ (upper frame) and  $q=1.1$ (lower frame).}
\label{fig:TsRates}
\end{figure}

In Fig.~\ref{fig:TsRates} deuterium reaction rates are plotted for
two different values of the $q$ parameter. We can observe that moderate
deviations $(q-1)\sim 0.1$ from the Maxwellian distribution increase
deuterium burning rates above the threshold rate $r_t
\approx 1 \textrm{sec}^{-1} \textrm{cm}^{-3}$: therefore, these processes
should contribute to heat Jupiter at the present epoch ($T\approx 1-2$ eV).
If we read the graphs for $T\approx 10-20$ eV, which corresponds to 
temperatures of the planet during its formation, we realize that it was
sufficient a smaller value, $(q-1)\sim 0.03$, to effect the thermal balance in
that period.

Reaction rates of the order of that required to heat Jupiter do not cause
decrease significantly the deuterium density inside the planet; in fact a
burning rate equal to ten times the threshold rate would consume a significant
fraction of deuterium only after a time of the order of $ n_D / (10 r_t)
\sim 10^{11}$ years.

These considerations demonstrate that deuterium burning is a possible
explanation for the Jupiter excess heat. Precise determinations of the
conditions inside Jupiter and additional microscopic calculations are
necessary for a better determination of the range of values of $q$
relevant to Jupiter interior.
%

%\vspace{-5mm}
\end{document}